# PERFORMANCE OF THE GOOGLE DESKTOP, ARABIC GOOGLE DESKTOP AND PEER TO PEER APPLICATION IN ARABIC LANGUAGE

Abd El Salam AL HAJJAR, Anis ISMAIL, Mohammad HAJJAR, Mazen EL-SAYED
University Institute of Technology
Lebanese University
Lebanon
abdsalamhajjar@hotmail.com
anismaiil@yahoo.com
m_hajjar@ul.edu.lb
mazen_elsayed@yahoo.fr

*ABSTRACT*

*The Arabic language is a complex language; it is different from Western languages especially at the morphological and spelling variations. Indeed, the performance of information retrieval systems in the Arabic language is still a problem. For this reason, we are interested in studying the performance of the most famous search engine, which is a Google Desktop, while searching in Arabic language documents. Then, we propose an update to the Google Desktop to take into consideration in search the Arabic words that have the same root. After that, we evaluate the performance of the Google Desktop in this context. Also, we are interested in evaluation the performance of peer-to-peer application in two ways. The first one uses a simple indexation that indexes Arabic documents without taking in consideration the root of words. The second way takes in consideration the roots in the indexation of Arabic documents. This evaluation is done by using a corpus of ten thousand documents and one hundred different queries.*

## 1. INTRODUCTION

The information retrieval based on multiple application and system such as: the search engine and peer-to-peer application. A search engine is communication software that allows finding resources which answer to a user request [1]. These resources can be web pages, images, videos, files, etc, which are represented by documents of different formats (HTML, JPEG, MPEG, PDF, etc.). The importance of this engine depends on relevance of the overall result that can contain million web pages. A peer-to-peer system allows to many computers to communicate over a network and sharing information's, files, continuous multimedia flows (streaming), a distributed computing, phones (such as Skype), etc.

The performance of the information retrieval systems varies with the used language, and depends on nature and complexity of the language, in which the request of research is formulated. These systems are mainly based on an automatic treatment of the natural language. These treatments change from one language to another, and may depend on particular characteristics of a language [2]. So, it is easy to see the role the structure of a natural language, in the way, in which one can access to the information in documents of the same language. The performance of search engines and peer-to-peer applications depends mainly on the efficiency of the indexing methods and the information retrieval, which constitute the heart of these systems [3] [4][5][6]. The powerful of the available search engines and peer-to-peer applications which are primarily developed for the Western languages, such as English, is increasing gradually. Although, it is clearly less, in case of the Arabic language, probably because of morphological specificities and structural characteristics of Arabic language compared to the Western languages [7][8][9][10][11][19][20]. Indeed, few studies have focused on studying the performance of such systems in Arabic language. For these reasons, we are interested in studying the performance of these engines and one peer-to-peer application to extract the relevant information from the Arabic documents. With this intention, we choose the most famous search engines, as Google, and we choose the version that can run on a local computer (Google Desktop), and we choose also one peer-to-peer application that we have developed. Then, we update the Google Desktop researcher by adding a layer that takes the query and finds its root and then retrieves all the words derived from this root, and submit the set of these all words to Google Desktop. In the other side, we update the indexation procedure in peer-to-peer application, for every word found in the document indexed with all the words derived from the same root. Therefore, we will present in this paper the performance of the Google Desktop, Google Desktop updated, peer-to-peer application with a simple indexation and peer-to-peer application with an advanced Arabic indexation in Arabic language.

The following section presents the general architecture of the Google search engine. In Section 3, we present the general architecture of the peer-to-peer application. In the section 4, we present the methodology and the corpus used to perform our experiments. Next, the results are given in Section 5. Finally, we finish by a conclusion.

## 2. SEARCH ENGINE

A search engine can provide a set of documents in response to a given query [1]. The entry of the engine is a query which can be only one word, a set of words or a phrase. The engine analyzes each word of the query and checks its index, while starting with the statistical analysis to find the documents containing exactly the word, or the phrase of the request. Then it tries to use the techniques of automatic processing of the natural language, to find a list of the most relevant documents. The result contains a short summary, containing the title and sometimes an outline of each document belonging to them. The search engines traverse all the visited pages of the web to feed their databases with copies of these documents. Then, the search engines analyze the contents of these documents, to determine the key words, as titles, headings, contexts of the document, etc. The resulting data are stored in a database [22] [23].

### 2.1. Google Desktop

Google Desktop is one of the most popular utility in desktop searches. It is designed for usage on a single-user Windows machine. In a multi-user environment, if user with administrative rights installs and runs Google Desktop, the index of files find by users, regardless of their owner. Google experienced negative publicity from a number of sources after the initial release of the product which has been widely reported in the press, with many cite as a potential security weakness. Just Google Desktop indexed all the files that access is given, highlighting the security issues of multi-user systems and the dependence of the administrative accounts on Windows, rather than the cause of these problems. For many, this represents a failure to design effective if is not secure.

Google Desktop also had other problems discovered in it, resulting from a study that is done by Rice University, indicating that the vulnerabilities existing in the integration of Google Desktop and the Google search engine on the Internet. Google has since claimed to have patched the vulnerabilities announced in this document, but did not discuss what steps have been taken to ensure this. Google has also maintained that there was no evidence to suggest that these vulnerabilities have been exploited (NA 2005 rapport).

The second release of the Google Desktop adds an improvement for user interface and the ability for users to determine what types of documents are initially indexed by the program - allowing users to have more control over files stored by the program. The second version of Google Desktop also added a "sidebar", an application that uses plug-ins to present information for both Internet and clean storage of Google Desktop. Plug-ins included pictures found on the computer, e-mail in recent years, weather information and a quick search [27].

## 2.1. Google Desktop Updated

In our study, we recall a pertinent document related to a query which is the document that contains the same query word or contains a word derived from the same root of the query word. For that, we update the Google Desktop researcher by adding a layer that takes the query and finds its root and retrieves all words derived from this root and submit the set of all these words to Google [23].

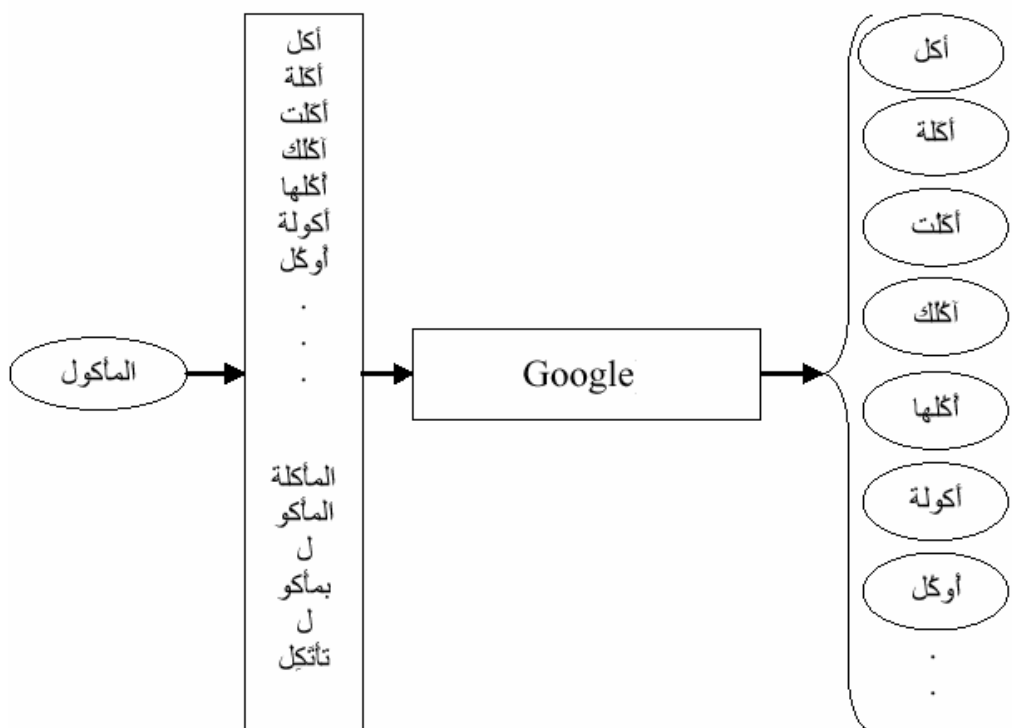

Figure 1. The documents must be found by updated Google engine in the case of query "المأكول" and its dependent words.

## 3. PEER-TO-PEER APPLICATION

The idea of peer-to-peer (P2P) computing offers new opportunities for building highly distributed data systems. Specifically, the P2P computing provides a very efficient way of storing and accessing the distributed resources. Peer-to-peer systems are distributed systems without any centralized control in which each node shares and exchanges data across the network (peer-to-peer network). The features of recent peer-to-peer systems: redundant storage, permanence, selection of nearby servers, anonymity, search, authentication, and hierarchical naming. They also offer the potential for low cost sharing of information, autonomy and privacy since they take the advantage of decentralization by distributing the storage information and computation cost among the peers, in addition to the ability to pool together and harness large amounts of resources. The strengths of existing P2P systems include self-organization, load-balancing, adaptation, and fault tolerance.

Before the appearance of internet access services by suppliers and the remarkable success of Napster [23], systems for sharing and exchanging information among computers were limited to client-server model such as the World Wide Web (WWW), local area networks (LAN) and software of FTP (File Transfer Protocol). Currently, the Internet is increasingly used; many applications use the network and consume bandwidth. Thus, the system has outgrown its original client-server design.

The peer-to-peer (P2P) systems search to form relations between the users for enabling them to pool resources such as processors, memory space, even if their initial motivation was to share files. They are used nowadays by various applications requiring decentralization. Paradigm (P2P) [25] began to flourish in a high growth by allowing each user of a network to play the role of client or server. In general, a P2P system is (more or less) composed (with or having) of a protocol for communication between peers. Algorithms finding the resources and application are at the top of the distributed environment, through direct exchange between peers. P2P technology allows an optimal sharing of computer resources and services such as information, files, processing and storage.

Napster systems [24] suggested downloading music files by using a central server for linking users. This allowed providing answers to queries in low delays. Then, the system Gnutella [26], fully decentralized, was implemented. Sharing information was so easy since any user could provide resources and get them on the network. Yet, the fact that these systems were decentralized posed another problem; i.e., how to get right answers to such queries while ensuring rapid and efficient way?

### 3.1. Indexation simple

P2P systems are widely used for sharing data or documents on a large scale. Usually, search query information, such as Google, is expressed by a set of keywords. In P2P systems, documents verifying these keywords (or part of these keywords) are considered relevant for this query. In contrast, in the domain of information retrieval, the goal is to get a list of the most relevant documents across the network. Thus, for the information retrieval in P2P systems, the challenge is not only to find the documents that are the most relevant to the user query, but also to retrieve documents efficiently. Our P2P system is a natural convergence between P2P systems and distributed databases.

Each peer shares data through relational database described by keywords. To find the relevant peers at this Query, this peer send its query to all its godfather "Super-Peer" do that matching keywords, describing the relations of the query with those described in its database and therefore these relevant relationships resent to the initiator peer.

## 3.2. Advanced Arabic indexation

P2P data indexing has recently attracted a great effort of many researches. For various proposed schemes, we enhance our method to operate with different queries from one keyword, like range of queries. When a peer sends a query with one word, we extract the root of this word then for each word derived set of word, each of it a single of query to be executed.

## 4. EVALUATION METHODOLOGY

We were interested to study the performance of the information retrieval systems for extracting relevant information from the Arabic documents. To do this, we selected the following research applications Google Desktop, Google Desktop updated, peer-to-peer application with simple indexation and peer-to-peer application with advanced Arabic indexation to perform our experiments. So we present in this part the performance of these applications in Arabic.

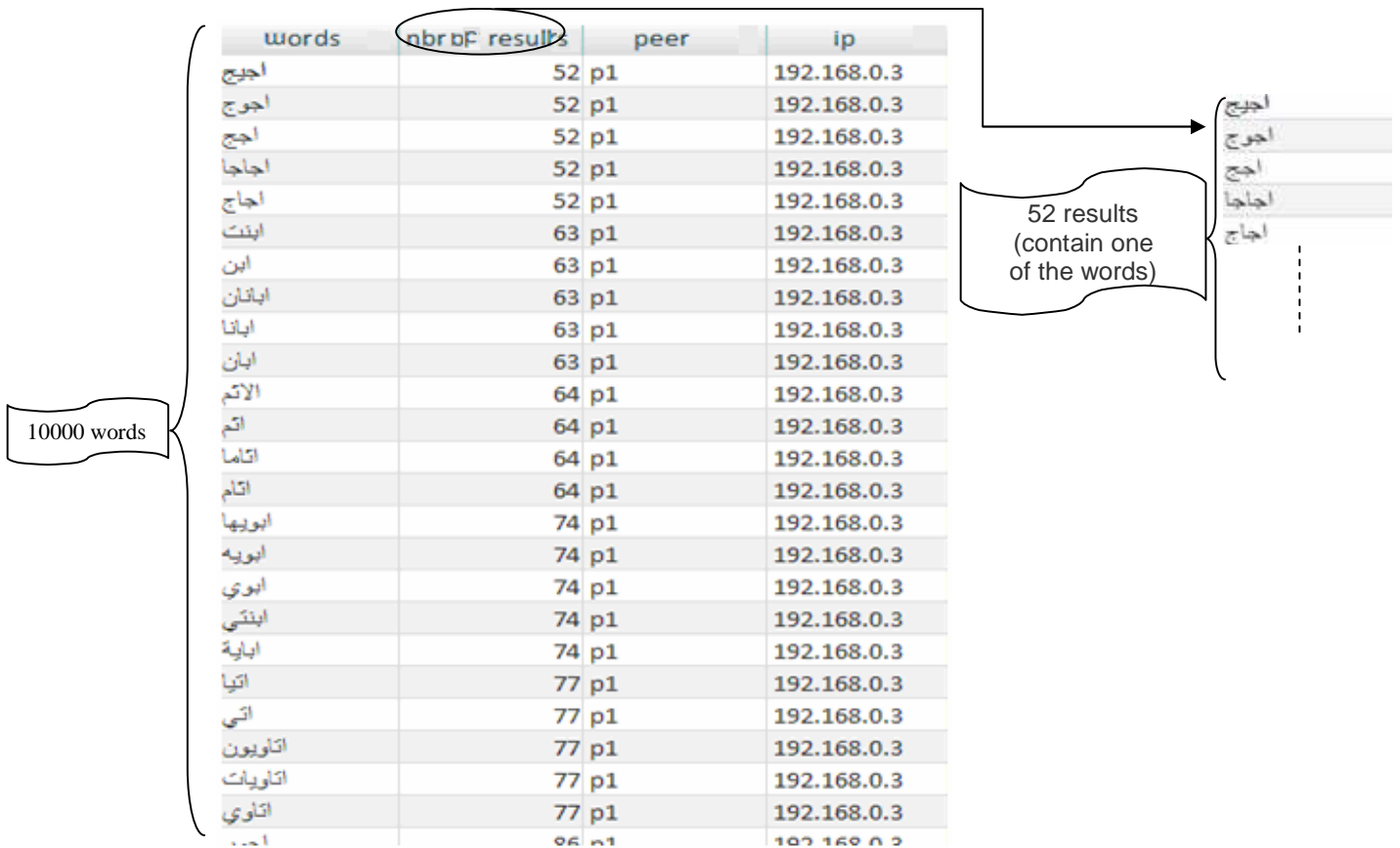

Figure 2. The words in Arabic with the probability of results for each word.

To evaluate the performance of research applications (search engines) on documents in Arabic, we installed research applications on network architecture in the peer. In addition, we have prepared an evaluation corpus with a set of queries and we followed an evaluation procedure to test or control our experiments [23].

### 4.1. Corpus

The corpus that we have built is a set of ten thousand documents in Arabic text format. The construction of this corpus is done in the following way:

- We selected 100 different Arabic roots (لعب, أكل,...).

- For each root, we selected 100 different words (يلعبون, يأكلون,.....).

- So we have 10,000 words from the words we generated, thus 10,000 documents each containing one of these words.

These documents are distributed on the peers of the network as follows:

- There are 4 peers; each peer contains documents that contain words that are related to 25 roots, which means we will have 2500 documents.

- As a result, there are two super peers, and then each super peer has 5000 documents that are related to 50 roots.

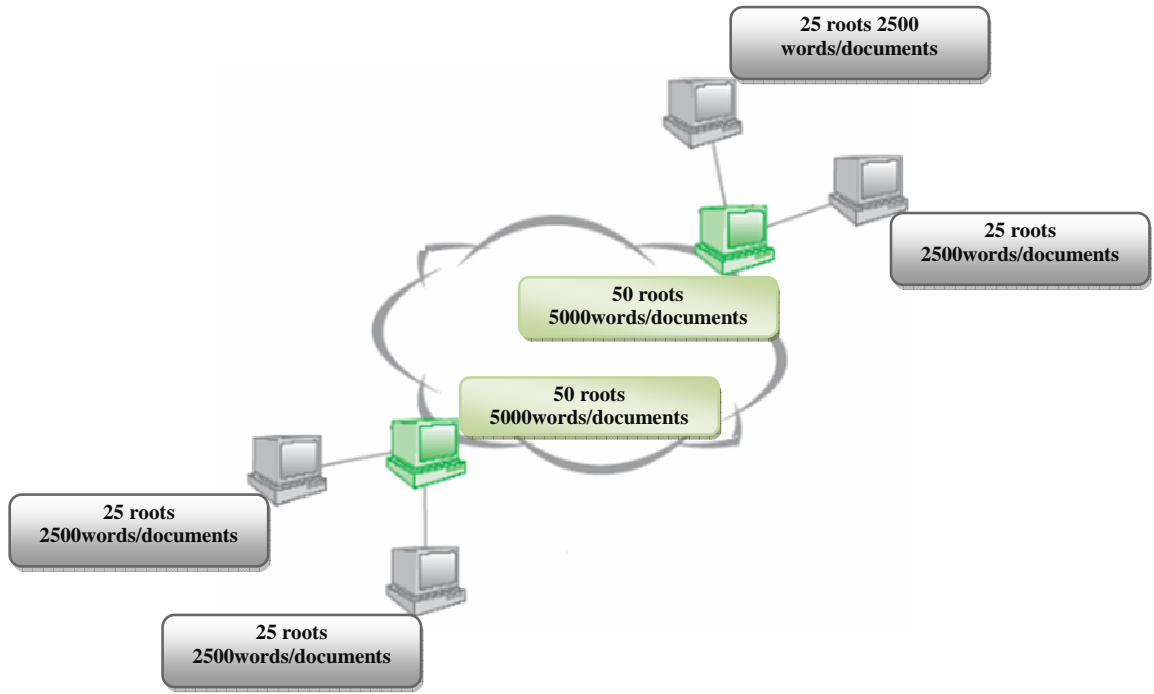

Figure 3. General architecture of Arab documents distribution on a peer-to-peer network.

## 4.2. Procedure

The procedure is done in an automatic way according to the following steps:

- For each application, we have implemented a function that takes as an input, a set of words (words as queries), then the user uses the procedure for each application, even Google desktop, because there is a publishing service for him, and finally this function is used to save queries with the results in a database.
- We chose a set of 100 queries, each query consists of a single word, and we have saved them in a database.
- We analyzed manually the relevant documents for each query, and attached the titles of these documents with each query
- Execute the 4 functions already implemented on the 100 requests
    1. For Google Desktop
    2. For Google Desktop Updated
    3. For the purpose peer-to-peer, which is the primary index for each document based on a single keyword (only the word that is in the document)
    4. For the purpose peer-to-peer, where the index is an advanced for each document according to several keywords (only the word that is in the document and all words that have the same roots).

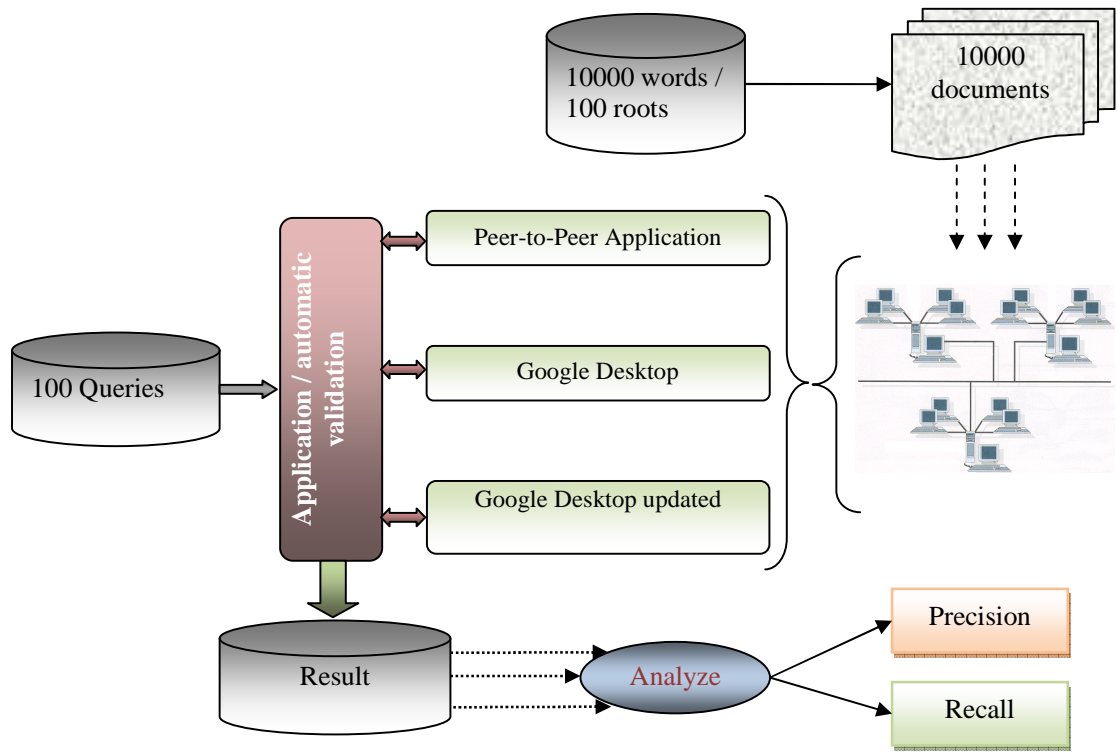

Figure 4. Procedure of research on peer to peer architecture with the three research applications.

## 4.2. Measures

To evaluate the results of each query, we used traditional measures, the precision and the recall that are used in information retrieval. Assuming that for a query Q, the SFound Results overview and SRelevant that is the number of relevant documents, then these measures are:

- Accuracy: For a query Q, the precision indicates the proportion of relevant documents among the documents found (1).

$$P = \frac{|S_{Relevant} \cap S_{Found}|}{S_{Found}} \quad (1)$$

- Reminder: For a query Q, recall measures the proportion of relevant documents in Q that have been found (2).

$$R = \frac{|S_{Relevant} \cap S_{Found}|}{|S_{Relevant}|} \quad (2)$$

## 5. RESULTS

As the evaluation procedure is done four times, so these results presented in four tables have the same structure. The first column gives the query started. These requests are all formed of a single word. The second column contains the key words of documents found for each query. The third and fourth columns present respectively the precision and recall for each query.

### 5.1. Google Desktop

Table 1. Results of hundred queries to Google Desktop.

| Query | Document contains | Precision | Recall | Query | Document contains | Precision | Recall |
|---|---|---|---|---|---|---|---|
| أَخُرْ | أَخُرْ | 1 | 0.0108 | أَصْلادٍ | أَصْلادٍ | 1 | 0.0161 |
| لَلْأَفِيكَةِ | لَلْأَفِيكَةِ | 1 | 0.0147 | يُضارِبُهُ | يُضارِبُهُ | 1 | 0.0063 |
| الأَئِمَّة | الأَئِمَّة | 1 | 0.0049 | يَطْبَعُه | يَطْبَعُه | 1 | 0.0100 |
| بحراني | بحراني | 1 | 0.0068 | الطِّرْماذِ | الطِّرْماذِ | 1 | 0.0625 |
| بَرْقُها | بَرْقُها | 1 | 0.0044 | طَهُرها | طَهُرها | 1 | 0.0073 |
| وبَصُرْتْ | وبَصُرْتْ | 1 | 0.0052 | ظهوركم | ظهوركم | 1 | 0.0030 |
| وابْتَكَرَ | وابْتَكَرَ | 1 | 0.0052 | والاسْتِعْتابُ | والاسْتِعْتابُ | 1 | 0.0063 |
| أبياتٌ | أبياتٌ | 1 | 0.0090 | والعَجَمات | والعَجَمات | 1 | 0.0052 |
| اثْرَدْتْ | اثْرَدْتْ | 1 | 0.0204 | عِذارُها | عِذارُها | 1 | 0.0027 |
| وثَوَرانَهُ | وثَوَرانَهُ | 1 | 0.0133 | وعِراض | وعِراض | 1 | 0.0017 |
| إجْذاعه | إجْذاعه | 1 | 0.0158 | يَعْرِقُ | يَعْرِقُ | 1 | 0.0033 |
| والجَزْمُ | والجَزْمُ | 1 | 0.0166 | عَشَوات | عَشَوات | 1 | 0.0096 |
| جمير | جمير | 1 | 0.0064 | بالمِعْضَد | بالمِعْضَد | 1 | 0.0068 |
| الجُهد | الجُهد | 1 | 0.0087 | عُقْدَةٌ | عُقْدَةٌ | 1 | 0.0049 |
| يَحْبِسُ | يَحْبِسُ | 1 | 0.0087 | عُمْرَه | عُمْرَه | 1 | 0.0033 |
| حُدود | حُدود | 1 | 0.00952 | عَنَنٌ | عَنَنٌ | 1 | 0.0088 |
| والحارِقَةُ | والحارِقَةُ | 1 | 0.00493 | تَغْييراً | تَغْييراً | 1 | 0.0069 |
| وحَسِرَ | وحَسِرَ | 1 | 0.00826 | غارِمٌ | غارِمٌ | 1 | 0.0137 |
| والمَخْضَرُ | والمَخْضَرُ | 1 | 0.00538 | وغَلَقَت | وغَلَقَت | 1 | 0.0086 |
| وحَكَمْتْ | وحَكَمْتْ | 1 | 0.00538 | الفُرُوج | الفُرُوج | 1 | 0.0069 |
| حَمَاةَ | حَمَاةَ | 1 | 0.01136 | بالفُرَّاط | بالفُرَّاط | 1 | 0.006 |
| نُحَمَّما | نُحَمَّما | 1 | 0.00503 | الفَطْرُ | الفَطْرُ | 1 | 0.0079 |
| خَبِيثٌ | خَبِيثٌ | 1 | 0.00926 | أَفْوَاقٍ | أَفْوَاقٍ | 1 | 0.0063 |
| وخَرْدَل | وخَرْدَل | 1 | 0.05882 | وقُدُوحٌ | وقُدُوحٌ | 1 | 0.0091 |
| المَخْصَرَة | المَخْصَرَة | 1 | 0.01176 | تَقْرِيراً | تَقْرِيراً | 1 | 0.0059 |
| كَخَفاه | كَخَفاه | 1 | 0.02128 | القَرَنِ | القَرَنِ | 1 | 0.0031 |
| وخَلِيْطِى | وخَلِيْطِى | 1 | 0.00599 | قَصْرُكَ | قَصْرُكَ | 1 | 0.0030 |
| والخَلِيقُ | والخَلِيقُ | 1 | 0.0030 | الإقْطَاع | الإقْطَاع | 1 | 0.0026 |
| المُخَيَّل | المُخَيَّل | 1 | 0.0069 | وقَلْبْتَ | وقَلْبْتَ | 1 | 0.0060 |
| والدَّرْدَرَةُ | والدَّرْدَرَةُ | 1 | 0.0142 | وقَنْطَرَ | وقَنْطَرَ | 1 | 0.0233 |
| أَدْهَمَهُ | أَدْهَمَهُ | 1 | 0.0108 | يُكَاتَبُ | يُكَاتَبُ | 1 | 0.0062 |
| ذَنائِبُ | ذَنائِبُ | 1 | 0.0065 | وتَكَرَّمَ | وتَكَرَّمَ | 1 | 0.0048 |
| وارْتَبَعوه | وارْتَبَعوه | 1 | 0.0021 | وتَكَلَّلَه | وتَكَلَّلَه | 1 | 0.0104 |
| الردف | الردف | 1 | 0.0076 | لَبَنَه | لَبَنَه | 1 | 0.0043 |
| وارْتَقَفُوا | وارْتَقَفُوا | 1 | 0.0060 | لِ۞قَحَتى | لِ۞قَحَتى | 1 | 0.0057 |
| رَمَلا | رَمَلا | 1 | 0.0082 | مَحقه | مَحقه | 1 | 0.0175 |
| وزابداً | وزابداً | 1 | 0.0117 | مُسْوك | مُسْوك | 1 | 0.0067 |
| وأَزْهَقَتْ | وأَزْهَقَتْ | 1 | 0.0128 | والمُلاَّحِيُّ | والمُلاَّحِيُّ | 1 | 0.0037 |
| المَسابل | المَسابل | 1 | 0.0080 | يَنْبُوعاً | يَنْبُوعاً | 1 | 0.0167 |
| سَريح | سَريح | 1 | 0.0068 | الناحِلُ | الناحِلُ | 1 | 0.0116 |
| يَسْفَحُه | يَسْفَحُه | 1 | 0.0172 | نسم | نسم | 1 | 0.0093 |
| والسَّلْسَلَة | والسَّلْسَلَة | 1 | 0.0153 | يَنْصِفُ | يَنْصِفُ | 1 | 0.0053 |
| سَمِعَ | سَمِعَ | 1 | 0.0041 | أَنفارٌ | أَنفارٌ | 1 | 0.0057 |
| وسَوَّدَتْ | وسَوَّدَتْ | 1 | 0.0050 | النَّواقِزِ | النَّواقِزِ | 1 | 0.0189 |
| بشرج | بشرج | 1 | 0.0093 | والمَنْهَرَةَ | والمَنْهَرَةَ | 1 | 0.0101 |
| الشُّعوبُ | الشُّعوبُ | 1 | 0.0080 | والهِرْماس | والهِرْماس | 1 | 0.0169 |
| اشمط | اشمط | 1 | 0.0158 | أَوْثَرَ | أَوْثَرَ | 1 | 0.0064 |
| صَباحٌ | صَباحٌ | 1 | 0.0048 | واسْتَوْرَدَه | واسْتَوْرَدَه | 1 | 0.0063 |
| صَراحة | صَراحة | 1 | 0.0102 | وضَعَتْ | وضَعَتْ | 1 | 0.0055 |
| الصَّعَافِقَةَ | الصَّعَافِقَةَ | 1 | 0.0434 | وَلَوْلْ | وَلَوْلْ | 1 | 0.0714 |
| **Toutes** | - | **100%** | **1.059%** | | | | |

## 5.2 Google Desktop updated

Table 2. Results of hundred queries to Google Desktop updated.

| Query | Document contains | Precision | Recall | Query | Document contains | Precision | Recall |
|---|---|---|---|---|---|---|---|
| أُخَرْ | أُخَرِ,... | 1 | 1 | أَصْلاد | أَصْلاد,... | 1 | 1 |
| لَلأَفِيكَةِ | لَلأَفِيكَةِ,... | 1 | 1 | يُضارِبُه | يُضارِبُه,... | 1 | 1 |
| الأَئِمَّة | الأَئِمَّة,... | 1 | 1 | يَطْبَعُه | يَطْبَعُه,... | 1 | 1 |
| بحرانى | بحرانى,... | 1 | 1 | الطِّرْماذِ | الطِّرْماذِ,... | 1 | 1 |
| بَرْقُها | بَرْقُها,... | 1 | 1 | ظَهْرها | ظَهْرها,... | 1 | 1 |
| وبَصُرْتُ | وبَصُرْتُ,... | 1 | 1 | ظهوركم | ظهوركم,... | 1 | 1 |
| وابْتَكَرَ | وابْتَكَرَ,... | 1 | 1 | والاسْتِعْتابُ | والاسْتِعْتابُ,... | 1 | 1 |
| أبياتٌ | أبياتٌ,... | 1 | 1 | والعَجَمات | والعَجَمات,... | 1 | 1 |
| اثْرَدْتُ | اثْرَدْتُ,... | 1 | 1 | عِذارُها | عِذارُها,... | 1 | 1 |
| وثَوَرانهُ | وثَوَرانهُ,... | 1 | 1 | وعِراض | وعِراض,... | 1 | 1 |
| إجْذاعه | إجْذاعه,... | 1 | 1 | يَعْرقُ | يَعْرقُ,... | 1 | 1 |
| والجَزْمُ | والجَزْمُ,... | 1 | 1 | عَشَوات | عَشَوات,... | 1 | 1 |
| جَمير | جَمير,... | 1 | 1 | بالمِعْضَد | بالمِعْضَد,... | 1 | 1 |
| الجُهد | الجُهد,... | 1 | 1 | عَقَدَةٌ | عَقَدَةٌ,... | 1 | 1 |
| يَحْبِسُ | يَحْبِسُ,... | 1 | 1 | عُمْرَه | عُمْرَه,... | 1 | 1 |
| حُدود | حُدود,... | 1 | 1 | عَنْ | عَنْ,... | 1 | 1 |
| والحارقة | والحارقة,... | 1 | 1 | تَغْييراً | تَغْييراً,... | 1 | 1 |
| وحَسِرَ | وحَسِرَ,... | 1 | 1 | غارِمٌ | غارِمٌ,... | 1 | 1 |
| والمَحْضَرُ | والمَحْضَرُ,... | 1 | 1 | وغَلِقَت | وغَلِقَت,... | 1 | 1 |
| وحَكَمْتُ | وحَكَمْتُ,... | 1 | 1 | الفُروج | الفُروج,... | 1 | 1 |

| Query | Document contains | Precision | Recall | | Query | Document contains | Precision | Recall |
|---|---|---|---|---|---|---|---|---|
| حَمَاةً | حَمَاةً... | 1 | 1 | | بِالفَرَّاطِ | بِالفَرَّاطِ... | 1 | 1 |
| تُحَمَّما | تُحَمَّما... | 1 | 1 | | الفُطْرُ | الفُطْرُ... | 1 | 1 |
| خَبِيثٍ | خَبِيثٍ... | 1 | 1 | | أفْوَاقٍ | أفْوَاقٍ... | 1 | 1 |
| وخَرْدَلٍ | وخَرْدَلٍ... | 1 | 1 | | وقُدُوحٍ | وقُدُوحٍ... | 1 | 1 |
| المَخْصَرَة | المَخْصَرَة... | 1 | 1 | | تَقْرِيراً | تَقْرِيراً... | 1 | 1 |
| كَخَفاه | كَخَفاه... | 1 | 1 | | القَرَنِ | القَرَنِ... | 1 | 1 |
| وخُلَيْطِي | وخُلَيْطِي... | 1 | 1 | | قَصْرُكِ | قَصْرُكِ... | 1 | 1 |
| والخَلِيقُ | والخَلِيقُ... | 1 | 1 | | الإقْطَاعِ | الإقْطَاعِ... | 1 | 1 |
| المُخَيَّلِ | المُخَيَّلِ... | 1 | 1 | | وقَلْبَتْ | وقَلْبَتْ... | 1 | 1 |
| والدَّرْدَرَةِ | والدَّرْدَرَةِ... | 1 | 1 | | وقَنْطَرِ | وقَنْطَرِ... | 1 | 1 |
| أدْهَمَهُ | أدْهَمَهُ... | 1 | 1 | | يُكَاتَبُ | يُكَاتَبُ... | 1 | 1 |
| ذَنائِبُ | ذَنائِبُ... | 1 | 1 | | وتَكَرَّمَ | وتَكَرَّمَ... | 1 | 1 |
| وارْتَبَعوه | وارْتَبَعوه... | 1 | 1 | | وتَكَلَّه | وتَكَلَّه... | 1 | 1 |
| الردف | الردف... | 1 | 1 | | لَبَنَه | لَبَنَه... | 1 | 1 |
| وارْتَقَفوا | وارْتَقَفوا... | 1 | 1 | | لَ◌فَحَتى | لَ◌فَحَتى... | 1 | 1 |
| رَمَلا | رَمَلا... | 1 | 1 | | مَحقَه | مَحقَه... | 1 | 1 |
| وزابداً | وزابداً... | 1 | 1 | | مُسُوك | مُسُوك... | 1 | 1 |
| وأزْهَقَتْ | وأزْهَقَتْ... | 1 | 1 | | والمُلاَّحِيُّ | والمُلاَّحِيُّ... | 1 | 1 |
| المَسَابِل | المَسَابِل... | 1 | 1 | | يَنْبُوعاً | يَنْبُوعاً... | 1 | 1 |
| سَرِيح | سَرِيح... | 1 | 1 | | الناحِلُ | الناحِلُ... | 1 | 1 |
| يَسْفَحُه | يَسْفَحُه... | 1 | 1 | | نسم | نسم... | 1 | 1 |
| والسَّلْسِلَةَ | والسَّلْسِلَةَ... | 1 | 1 | | يَنْصَِفُ | يَنْصَِفُ... | 1 | 1 |
| سَمِعَ | سَمِعَ... | 1 | 1 | | أنْفارٍ | أنْفارٍ... | 1 | 1 |
| وسَوَّدْتَ | وسَوَّدْتَ... | 1 | 1 | | النَواقِزِ | النَواقِزِ... | 1 | 1 |
| بشرج | بشرج... | 1 | 1 | | والمَنْهَرَةِ | والمَنْهَرَةِ... | 1 | 1 |
| الشَّعوبُ | الشَّعوبُ... | 1 | 1 | | والهِرْماسُ | والهِرْماسُ... | 1 | 1 |
| اشمط | اشمط... | 1 | 1 | | أوْتَرَ | أوْتَرَ... | 1 | 1 |
| صِباحٌ | صِباحٌ... | 1 | 1 | | واسْتَوْرَدَه | واسْتَوْرَدَه... | 1 | 1 |
| صَراحة | صَراحة... | 1 | 1 | | وضَعَتْ | وضَعَتْ... | 1 | 1 |
| الصَعَافِقَةُ | الصَعَافِقَةُ... | 1 | 1 | | وَلَوْلٌ | وَلَوْلٌ... | 1 | 1 |
| **Toutes** | **-** | **100%** | **100%** | | | | | |

### 5.3. Peer-to-Peer application with simple indexation

Table 3. Results of hundred queries to Peer-to-Peer application with simple indexation

| Query | Document contains | Precision | Recall | | Query | Document contains | Precision | Recall |
|---|---|---|---|---|---|---|---|---|
| أخْرَ | أخْرَ | 1 | 0.0108 | | أصْلاد | أصْلاد | 1 | 0.0161 |
| لِلأفِيكَة | لِلأفِيكَة | 1 | 0.0147 | | يُضارِبُه | يُضارِبُه | 1 | 0.0063 |
| الأئِمَّة | الأئِمَّة | 1 | 0.0049 | | يَطْبَعُه | يَطْبَعُه | 1 | 0.0100 |
| بحرانى | بحرانى | 1 | 0.0068 | | الطِرْماذ | الطِرْماذ | 1 | 0.0625 |
| بَرْقُها | بَرْقُها | 1 | 0.0044 | | طُهْرُها | طُهْرُها | 1 | 0.0073 |
| وبَصُرْتُ | وبَصُرْتُ | 1 | 0.0052 | | ظهورِكم | ظهورِكم | 1 | 0.0030 |
| وابْتَكَرَ | وابْتَكَرَ | 1 | 0.0052 | | والاسْتِعْنابُ | والاسْتِعْنابُ | 1 | 0.0063 |
| أبياتٌ | أبياتٌ | 1 | 0.0090 | | والعَجَمات | والعَجَمات | 1 | 0.0052 |
| اثْرَدْتُ | اثْرَدْتُ | 1 | 0.0204 | | عِذارُها | عِذارُها | 1 | 0.0027 |
| وثُؤْرانه | وثُؤْرانه | 1 | 0.0133 | | وعِراض | وعِراض | 1 | 0.0017 |
| إجْذاعه | إجْذاعه | 1 | 0.0158 | | يَعْرَقُ | يَعْرَقُ | 1 | 0.0033 |
| والجَزْمُ | والجَزْمُ | 1 | 0.0166 | | عَشَواتٍ | عَشَواتٍ | 1 | 0.0096 |
| جَمِير | جَمِير | 1 | 0.0064 | | بالمِعْضَد | بالمِعْضَد | 1 | 0.0068 |
| الجُهد | الجُهد | 1 | 0.0087 | | عُقْدَة | عُقْدَة | 1 | 0.0049 |
| يَحْبِسُ | يَحْبِسُ | 1 | 0.0087 | | عُمْرَه | عُمْرَه | 1 | 0.0033 |
| حُدود | حُدود | 1 | 0.00952 | | عَنْنُ | عَنْنُ | 1 | 0.0088 |
| والحارقَة | والحارقَة | 1 | 0.00493 | | تَغْبِيراً | تَغْبِيراً | 1 | 0.0069 |
| وحَبِسَ | وحَبِسَ | 1 | 0.00826 | | غارِمٌ | غارِمٌ | 1 | 0.0137 |
| والمَحْضَرُ | والمَحْضَرُ | 1 | 0.00538 | | وغَلِقَت | وغَلِقَت | 1 | 0.0086 |
| وحَكَمْتْ | وحَكَمْتْ | 1 | 0.00538 | | الفُرُوج | الفُرُوج | 1 | 0.0069 |
| حَماة | حَماة | 1 | 0.01136 | | بالفَرَّاط | بالفَرَّاط | 1 | 0.006 |
| تُحَمَّما | تُحَمَّما | 1 | 0.00503 | | الفُطْرُ | الفُطْرُ | 1 | 0.0079 |
| خَبِيثٌ | خَبِيثٌ | 1 | 0.00926 | | أفْوَاقٍ | أفْوَاقٍ | 1 | 0.0063 |
| وخَرْدَلَ | وخَرْدَلَ | 1 | 0.05882 | | وقُدُوحٌ | وقُدُوحٌ | 1 | 0.0091 |
| المَخْصَرَة | المَخْصَرَة | 1 | 0.01176 | | تَقْرِيراً | تَقْرِيراً | 1 | 0.0059 |
| كَخَفاه | كَخَفاه | 1 | 0.02128 | | القَرَنِ | القَرَنِ | 1 | 0.0031 |
| وخَلْيْطى | وخَلْيْطى | 1 | 0.00599 | | قَصْرُك | قَصْرُك | 1 | 0.0030 |
| والخَلِيقُ | والخَلِيقُ | 1 | 0.0030 | | الإقْطَاع | الإقْطَاع | 1 | 0.0026 |
| المُخَيَّلِ | المُخَيَّلِ | 1 | 0.0069 | | وقَلْبْتْ | وقَلْبْتْ | 1 | 0.0060 |
| والدَّرْدَرَةُ | والدَّرْدَرَةُ | 1 | 0.0142 | | وقَنْطَر | وقَنْطَر | 1 | 0.0233 |
| أدْهَمَهُ | أدْهَمَهُ | 1 | 0.0108 | | يُكاتَبُ | يُكاتَبُ | 1 | 0.0062 |
| ذَنائبُ | ذَنائبُ | 1 | 0.0065 | | وتَكَرَّمَ | وتَكَرَّمَ | 1 | 0.0048 |
| وارْتَبَعوه | وارْتَبَعوه | 1 | 0.0021 | | وتَكَلَّه | وتَكَلَّه | 1 | 0.0104 |
| الردف | الردف | 1 | 0.0076 | | لَبَنه | لَبَنه | 1 | 0.0043 |
| وارْتَقْفوا | وارْتَقْفوا | 1 | 0.0060 | | لِ◌ قْحَتى | لِ◌ قْحَتى | 1 | 0.0057 |
| رَمَلا | رَمَلا | 1 | 0.0082 | | مَحقه | مَحقه | 1 | 0.0175 |
| وزَابداً | وزَابداً | 1 | 0.0117 | | مُسْوك | مُسْوك | 1 | 0.0067 |
| وأزْهَقْتْ | وأزْهَقْتْ | 1 | 0.0128 | | والمُلاَّجِيُّ | والمُلاَّجِيُّ | 1 | 0.0037 |
| المَسابل | المَسابل | 1 | 0.0080 | | يَنْبوعاً | يَنْبوعاً | 1 | 0.0167 |
| سَريح | سَريح | 1 | 0.0068 | | الناجلُ | الناجلُ | 1 | 0.0116 |
| يَسْفَحُه | يَسْفَحُه | 1 | 0.0172 | | نسم | نسم | 1 | 0.0093 |
| والسَّلْسلة | والسَّلْسلة | 1 | 0.0153 | | يَنْصِفُ | يَنْصِفُ | 1 | 0.0053 |
| سَمِعَ | سَمِعَ | 1 | 0.0041 | | أنْفارٌ | أنْفارٌ | 1 | 0.0057 |
| وسَوَّدْتْ | وسَوَّدْتْ | 1 | 0.0050 | | النَّواقِز | النَّواقِز | 1 | 0.0189 |
| بشرج | بشرج | 1 | 0.0093 | | والمَنْهَرَةُ | والمَنْهَرَةُ | 1 | 0.0101 |
| الشَّعوبُ | الشَّعوبُ | 1 | 0.0080 | | والهِرْماسُ | والهِرْماسُ | 1 | 0.0169 |
| اشْمَطْ | اشْمَطْ | 1 | 0.0158 | | أوْثَرَ | أوْثَرَ | 1 | 0.0064 |
| صباحٌ | صباحٌ | 1 | 0.0048 | | واسْتَوْرَدَه | واسْتَوْرَدَه | 1 | 0.0063 |
| صَراحة | صَراحة | 1 | 0.0102 | | وضعَتْ | وضعَتْ | 1 | 0.0055 |
| الصَّعَافقَة | الصَّعَافقَة | 1 | 0.0434 | | ولَوْلْ | ولَوْلْ | 1 | 0.0714 |
| **Toutes** | **-** | **100%** | **1.059%** | | | | | |

## 5.4. Peer-to-Per application with advanced Arabic indexation

Table 4. Results of hundred queries to peer-to-peer application with advanced Arabic indexation

| Query | Document contains | Precision | Recall | | Query | Document contains | Precision | Recall |
|---|---|---|---|---|---|---|---|---|
| أخْرْ | أخْرْ... | 1 | 1 | | أصْلاد | أصْلاد... | 1 | 1 |
| للأفيكة | للأفيكة... | 1 | 1 | | يُضارِبُه | يُضارِبُه... | 1 | 1 |
| الأئَمَّة | الأئَمَّة... | 1 | 1 | | يَطْبَعُه | يَطْبَعُه... | 1 | 1 |
| بحراني | بحراني... | 1 | 1 | | الطُرْماذ | الطُرْماذ... | 1 | 1 |
| بَرْقُها | بَرْقُها... | 1 | 1 | | طُهْرها | طُهْرها... | 1 | 1 |
| وبَصُرْتْ | وبَصُرْتْ... | 1 | 1 | | ظهوركم | ظهوركم... | 1 | 1 |
| وابْتَكَرَ | وابْتَكَرَ... | 1 | 1 | | والاسْتَعْتابُ | والاسْتَعْتابُ... | 1 | 1 |
| أبياتٌ | أبياتٌ... | 1 | 1 | | والعَجَمات | والعَجَمات... | 1 | 1 |
| اثْرَدْتْ | اثْرَدْتْ... | 1 | 1 | | عِذارُها | عِذارُها... | 1 | 1 |
| وثَوَرانُه | وثَوَرانُه... | 1 | 1 | | وعِراض | وعِراض... | 1 | 1 |
| إجذاعه | إجذاعه... | 1 | 1 | | يَعْرِقُ | يَعْرِقُ... | 1 | 1 |
| والجَزْمُ | والجَزْمُ... | 1 | 1 | | عَشْوات | عَشْوات... | 1 | 1 |
| جَمير | جَمير... | 1 | 1 | | بالمِعْضَد | بالمِعْضَد... | 1 | 1 |
| الجُهد | الجُهد... | 1 | 1 | | عُقْدَةٌ | عُقْدَةٌ... | 1 | 1 |
| يَحْبِسُ | يَحْبِسُ... | 1 | 1 | | عُمْرَه | عُمْرَه... | 1 | 1 |
| حُدُود | حُدُود... | 1 | 1 | | عَنْ | عَنْ... | 1 | 1 |
| والحارقةُ | والحارقةُ... | 1 | 1 | | تَغْييراً | تَغْييراً... | 1 | 1 |

| | | | | | | | |
|---|---|---|---|---|---|---|---|
| وحَسِرَ | وحَسِرَ,... | 1 | 1 | غارِمٌ | غارِمٌ,... | 1 | 1 |
| والمَحْضَرُ | والمَحْضَرُ,... | 1 | 1 | وغَلَقَتْ | وغَلَقَتْ,... | 1 | 1 |
| وحَكَمْتُ | وحَكَمْتُ,... | 1 | 1 | الفُرُوج | الفُرُوج,... | 1 | 1 |
| حَمَاةَ | حَمَاةَ,... | 1 | 1 | بالفِرَّاطِ | بالفِرَّاطِ,... | 1 | 1 |
| تَحَمَّما | تَحَمَّما,... | 1 | 1 | الفُطْرُ | الفُطْرُ,... | 1 | 1 |
| خَبِيثٌ | خَبِيثٌ,... | 1 | 1 | أَفْوَاقٌ | أَفْوَاقٌ,... | 1 | 1 |
| وخَرْدَلَ | وخَرْدَلَ,... | 1 | 1 | وقُدُوحٌ | وقُدُوحٌ,... | 1 | 1 |
| المَخْصَرَةُ | المَخْصَرَةُ,... | 1 | 1 | تَقْرِيراً | تَقْرِيراً,... | 1 | 1 |
| كَخَفَاه | كَخَفَاه,... | 1 | 1 | القَرَن | القَرَن,... | 1 | 1 |
| وخَلِيطِي | وخَلِيطِي,... | 1 | 1 | قَصْرُكَ | قَصْرُكَ,... | 1 | 1 |
| والخَلِيقُ | والخَلِيقُ,... | 1 | 1 | الإِقْطَاع | الإِقْطَاع,... | 1 | 1 |
| المُخَيَّل | المُخَيَّل,... | 1 | 1 | وقَلَبْتُ | وقَلَبْتُ,... | 1 | 1 |
| والدَّرْدَرَةُ | والدَّرْدَرَةُ,... | 1 | 1 | وقَنْطَرَ | وقَنْطَرَ,... | 1 | 1 |
| أَدْهَمَهُ | أَدْهَمَهُ,... | 1 | 1 | يُكاتَبُ | يُكاتَبُ,... | 1 | 1 |
| ذَنائِبُ | ذَنائِبُ,... | 1 | 1 | وتَكَرَّمَ | وتَكَرَّمَ,... | 1 | 1 |
| وارْتَبَعوه | وارْتَبَعوه,... | 1 | 1 | وتَكَلَّله | وتَكَلَّله,... | 1 | 1 |
| الردف | الردف,... | 1 | 1 | لَبَنَه | لَبَنَه,... | 1 | 1 |
| وارْتَقَفوا | وارْتَقَفوا,... | 1 | 1 | لِ⊙فَحَتى | لِ⊙فَحَتى,... | 1 | 1 |
| رَمَلاً | رَمَلاً,... | 1 | 1 | مَحقّه | مَحقّه,... | 1 | 1 |
| وزابداً | وزابداً,... | 1 | 1 | مُسُوك | مُسُوك,... | 1 | 1 |
| وأَزْهَقَتْ | وأَزْهَقَتْ,... | 1 | 1 | والمُلَاحِيُّ | والمُلَاحِيُّ,... | 1 | 1 |
| المَسابِل | المَسابِل,... | 1 | 1 | يَنْبُوعاً | يَنْبُوعاً,... | 1 | 1 |
| سَرِيح | سَرِيح,... | 1 | 1 | الناحِلُ | الناحِلُ,... | 1 | 1 |
| يَسْفَحُه | يَسْفَحُه,... | 1 | 1 | نسم | نسم,... | 1 | 1 |
| والسَّلْسِلَةُ | والسَّلْسِلَةُ,... | 1 | 1 | يَنْصَفُ | يَنْصَفُ,... | 1 | 1 |
| سَمِعَ | سَمِعَ,... | 1 | 1 | أَنفارٌ | أَنفارٌ,... | 1 | 1 |
| وسَوَّدَتْ | وسَوَّدَتْ,... | 1 | 1 | النَّواقِزُ | النَّواقِزُ,... | 1 | 1 |
| بِشْرِج | بِشْرِج,... | 1 | 1 | والمَنْهَرَةُ | والمَنْهَرَةُ,... | 1 | 1 |
| الشَّعوبُ | الشَّعوبُ,... | 1 | 1 | والهِرْماسُ | والهِرْماسُ,... | 1 | 1 |
| اشمَطَّ | اشمَطَّ,... | 1 | 1 | أَوْتَرَ | أَوْتَرَ,... | 1 | 1 |
| صِباحٌ | صِباحٌ,... | 1 | 1 | واسْتَوْرَدَه | واسْتَوْرَدَه,... | 1 | 1 |
| صَراحة | صَراحة,... | 1 | 1 | وضَعَتْ | وضَعَتْ,... | 1 | 1 |
| الصَّعَافِقَةُ | الصَّعَافِقَةُ,... | 1 | 1 | وَلْوَلْ | وَلْوَلْ,... | 1 | 1 |
| **Toutes** | - | **100%** | **100%** | | | | |

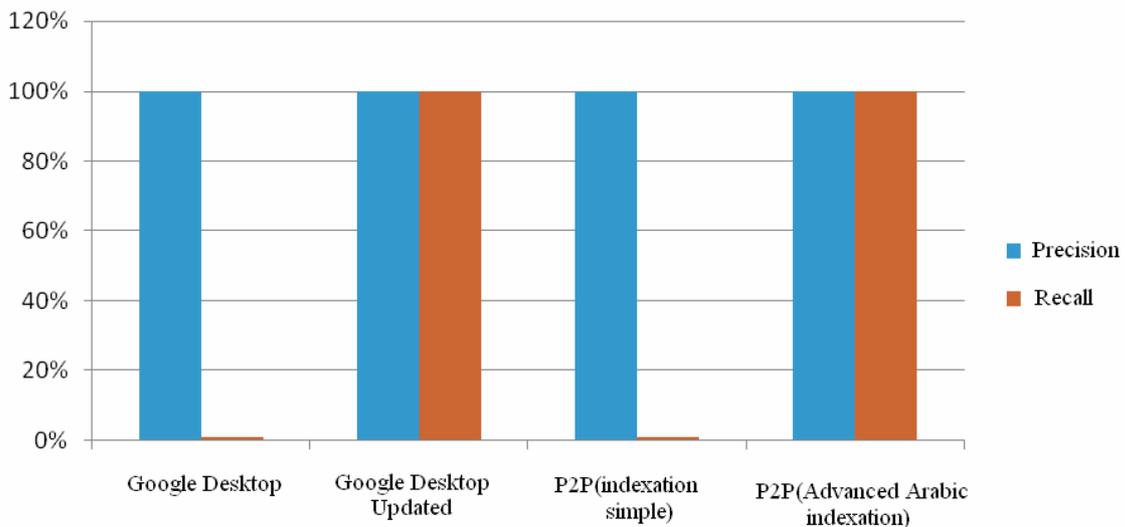

Figure 4. Successful percentage for every application.

Since the functions implemented for each search application take as an input a set of words (words as queries), the evaluation operation is simple and fast. It provides the performance of search application, which differs from an application to another. The results of hundreds of queries to the Google Desktop are presented in table1. These results show that the Google Desktop that can retrieve documents, contain exactly the query word and whatever the word is used. For this reason, the precision of the word used in Google Desktop, it is 1 (one document found) or 0 (no documents found). Similarly, the recall of the word, it is 5% (one document found among twenty relevant) or 0% (no documents found). For example, if we have a query "أُخْر" Google Desktop can retrieve only the documents containing the same form of the word, without changing any letter of the word. Thus, Google Desktop can retrieve the document containing others words which are derived from the same root of "أُخْر". So it seems that the Google Desktop considers theform of a written word without analyzing it. In addition, the average precision of the Google Desktop is 100%, because Google Desktop retrieves at least one document that contains the same query word and this document is usually a relevant document and the average recall is about 1%. From that result, the problems of the Google Desktop appear in its local version, in the extraction of information from Arabic documents. It seems that the specific treatments for Arabic language, particularly the morphological analysis, are not included in this search engine.

After the results of Google Desktop, we take the experiments that we performed by changing the query to add the words of the same group (have same root) to the keyword. This is achieved through an application that sits between the user query and the Google Desktop to obtain the Google Desktop Updated. We obtain the results of hundreds of queries to the Google Desktop Updated are presented in table2. We reach the value of 100% for both precision and recall. In consideration that the relevant documents for a query are the documents contain a word have the same root of the initial query word.

Similarly, Peer-to-Peer application with simple indexation results is present in the table 3. These results show also that this application that can retrieve documents, contain exactly the query word and whatever the word is used. To achieve these results we repeat this evaluation on Peer-to-Peer application with advanced Arabic indexation, as the Google Desktop Updated, we reach the value of 100% for both precision and recall.

## 6. CONCLUSION

In this paper, we are interested in studying the performance of the search engine Google Desktop on documents in Arabic language. Also, we have proposed an update to the Google Desktop that uses the techniques of root extraction in Arabic language in order to increase the performance of Google Desktop when the request concerns Arabic documents, and we have evaluated the performance of this engine in this context. We are interested also to evaluate the performance of Peer-to-Peer application in two ways. The first one uses a simple indexation that indexes Arabic documents without taking in consideration the root of words. The second way takes in consideration the roots in the indexation of Arabic documents. The results obtained in the previous section show clearly that the use of Arabic root extraction improves clearly the result of research on Arabic documents.

## ACKNOWLEDGEMENTS

This work has been done as a part of the projects "Automatic information extraction form Arabic texts" by CNRSL, "Extraction automatique d'information à partir des documents Arabes", UL and

"Arabic speech synthesis from text, with natural prosody, using linguistic and semantic analysis" PCSL.

**Authors**

Dr. Abd El Salam AL HAJJAR, Born in Lebanon, work as instructor at the Lebanese University, University Institute of technology, Sidon, Lebanon. He has a B.S and Technical leader at the oger system company, Lebanon Branch. He has a B.A in applied Mathematics, Computer Science from the Lebanese University – Faculty of Sciences, and Masters in Computer Science "Cooperation in sciences of information treatment" from the Lebanese university and Paul Sabatier University (IRIT France), and a Ph.D. in Computer Science from Paris8 University, France. His main research in the Arabic information extraction and processing.

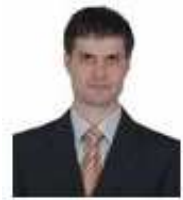

Dr. Anis Ismail, Born in Lebanon, works as system and network administrator and instructor at the Lebanese University, University Institute of technology, Sidon, Lebanon. He has a B.S. degree in Telecommunication and Networking Engineering from the Lebanese University (LU), an M.S. in Computer Science and MS CCE from the American University of Science and Technology (AUST) in Lebanon, and a Ph.D. in Computer Science from the University of AIX-Marseille, France. His main research interest covers Data Mining in P2P Systems, Arabic Language Processing, and Multimedia Information.

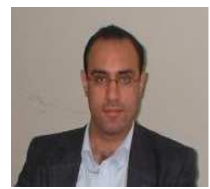

Dr. Mohammad Hajjar is a Professor at University Institute of Technology, Lebanese University, in Lebanon. He received a PHD in computer Science at Nantes University in France. His Interest domain concerns Arabic language processing, multimedia information research and data management in peer-to- peer systems.

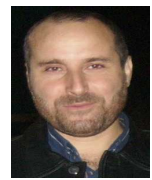

Dr. Mazen EL-SAYED, Born in Lebanon, works as assistance professor, and Head of Applied Bussiness Computer Department, at the Lebanese University, University Institute of technology, Sidon, Lebanon. He has an engineer degree in computer science from the Lebanese University (LU), an M.S. in Computer Science from the Central School of Engineering (ECN), University of Nantes, France, and a Ph.D. in Computer Science from the Anger University, France.

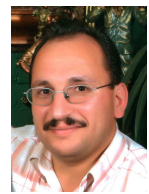